# Graham's Schedules and the Number Partition Problem (NPP)

Seenu S. Reddi
ReddiSS at aol dot com
July 19, 2008

Abstract: We show the equivalence of the Number Partition Problem and the two processor scheduling problem. We establish a priori bounds on the completion times for the scheduling problem which are tighter than Graham's but almost on par with a posteriori bounds of Coffman and Sethi. We conclude the paper with a characterization of the asymptotic behavior of the scheduling problem which relates to the spread of the processing times and the number of jobs.

There is recent interest in the Number Partition Problem (NPP) (for instance see [1], [2]) and solutions are presented based on the Karp/Karmarkar technique for optimal partitions. We do not concern ourselves with this approach but rather solve the problem by pointing out the equivalence between the NPP and 2-Processor Scheduling (2PS), and using Graham's schedules. Graham's schedules are usually called Longest Processing Time (LPT) schedules in the literature but in view of his pioneering work in discovering these almost optimal schedules for the multi-processing problem and the fact that this discovery ranks next to the Selmer Johnson's beautiful result about the two-machine flow-shop scheduling, we feel justified in our nomenclature. In the course of our solution, we present a priori bounds tighter than Graham's and a posteriori bounds comparable to Coffman and Sethi [3]. We also derive asymptotic bounds when the numbers get large for specific conditions.

We will state the two problems NPP and 2PS and establish the equivalence by a simple argument. Though this equivalence seems to be known and suspected by workers in the field, there was no simple and explicit proof offered (to the knowledge of the author). Assume a set of n numbers $K = \{t_1, t_2, \ldots, t_n\}$ which are positive and greater than zero. Define $S(K) = t_1 + t_2 + \ldots, + t_n$ as the sum of the members of the set K.

**NPP:** Partition the set K into two sets $K_1$ and $K_2$ so that the absolute difference between $S(K_1)$ and $S(K_2)$ is minimal.

**2PS:** Schedule the independent tasks with processing times $t_1, t_2, \ldots, t_n$ on two identical processors with identical speeds so that the completion time is minimized.

Note that 2PS minimizes the completion time (also referred to as make span) whereas the NPP minimizes the idle time on the processors.

**Assertion:** NPP $\Leftrightarrow$ 2PS.
Proof: Assume the exclusive subsets $A_1$ and $A_2$ of K is an optimal solution to the NPP and $B_1$ and $B_2$ is an optimal solution to the 2PS. Assume for simplicity $S(A_1) \geq S(A_2)$ and $S(B_1) \geq S(B_2)$. Then it follows since $\{B_1, B_2\}$ minimizes the completion time and $\{A_1, A_2\}$ minimizes the idle time:

$$S(A_1) \geq S(B_1) \tag{1}$$
$$S(B_1) - S(B_2) \geq S(A_1) - S(A_2) \tag{2}$$

From these two relations we have:

$$S(B_1) + (S(A_2) - S(B_2)) \geq S(A_1) \geq S(B_1) \tag{3}$$

Since $S(A_1) + S(A_2) = S(B_1) + S(B_2)$, we have from (1):

$$S(A_2) - S(B_2) = S(B_1) - S(A_1) \leq 0 \tag{4}$$

From (3) we have $(S(A_2) - S(B_2)) \geq 0$ and combining with (4), it follows:

$$S(A_2) - S(B_2) = S(B_1) - S(A_1) = 0 \tag{5}$$

This proves the assertion that the two solutions are equivalent.

Since the NPP is equivalent to the 2PS, hereafter we will exclusively concern ourselves with the two processor scheduling problems and its solutions. There is extensive literature on this problem and Chen has a very comprehensive summary in his article [4]. Also we consider only two processors in the sequel and the number of processors is fixed at two. Most of the results can be extended to multiple processors but we do not do so here.

We assume the set of jobs $K = \{t_1, t_2, \ldots, t_n\}$ with $t_1 \geq t_2 \geq \ldots \geq t_n$. Graham's schedule (or LPT schedule) is to schedule the task with the longest processing time on one of the two processors when it becomes available. Thus a set of jobs with processing times {9, 7, 4, 3, 2} will be completed with a time of 13 units – processor 1 will be assigned tasks with processing times 9 and 3, whereas processor 2 with tasks with processing times 7, 4 and 2. Let $C_O$ be the optimal completion time and $C_G$ the completion time for Graham's schedule. Then Graham proved [5]:

**Theorem 1:** $C_G / C_O \leq 7 / 6$ for two processors.
Proof: See [5] for the proof.

Graham's bound is a priori in the sense that the bound is given without consideration of the jobs involved. Coffman and Sethi [3] provide a better, a posteriori bound based on the number of jobs scheduled on the processor that finishes the last. In the above example, processor 2 finishes last with three jobs and a processing time of 13. Let k be the number of jobs assigned to the processor that finishes the last.

**Theorem 2:** $C_G / C_O \leq 1 + 1/k - 1/2k$ for two processors.
Proof: Proof is involved and the interested reader may refer to [3]. Bo Chen [6] has a correction but is not applicable to the two processor case.

We will now derive an improved bound better than Graham's in a relatively simple fashion based on the last job scheduled on the two processors. We assume the number of jobs n >> 2 and name the jobs with processing times $t_1, t_2, \ldots, t_n$ as $J_1, J_2, \ldots, J_n$ respectively. Let $J_L$ be the last job to be finished in the Graham's schedule. In the above example $J_L$ will be $J_5$ since the last job to be finished has a processing time of two units and there are five jobs $J_1, J_2, \ldots, J_5$ with processing times 9, 7, .., 2 respectively. Let L be the index (or subscript) of the last job to be finished and $M = \lceil L / 2 \rceil$, i.e., M is the least integer $\geq L / 2$.

**Theorem 3:** $C_G / C_O \leq (P + 1) / P - 1 / 2P$ where $P = 24M^3 / (7 + 12M + 24M^2)$.
Proof: We will find a P such that $C_G / C_O \leq (P + 1) / P - 1 / 2P$. To find such a P, first find a P that violates the bound $C_G / C_O > (P + 1) / P - 1 / 2P$. Following Graham's proof, we get:

$$t_L > C_0 / P \tag{6}$$

Since $C_G / C_O \leq 7 / 6$ from Theorem 1, we have:

$$C_O \geq 6C_G / 7 \tag{7}$$

Combining (6) and (7) we get:

$$t_L > 6C_G / (7P) \tag{8}$$

Thus we should choose P such that (8) is violated to get the required P:

$$t_L \leq 6C_G / (7P) \tag{9}$$

Since $Mt_L \leq C_G$, it follows:

$$1 \leq 6M / (7P) \tag{10}$$

We can select P to be:

$$P = 6M / 7 \tag{11}$$

We can iterate this approach once more but starting with a better bound:

$$C_G / C_O \leq (P + 1) / P - 1 / 2P \text{ where } P = (12M + 7) / 12M \tag{12}$$

obtained from (11).

Iterating this procedure a couple of times, we get Theorem 3.

The reader may object to the proof in the sense it is not direct, i.e., assuming violation and then establishing the right value for the bounds. The problem is that if we do not do it, we get $t_L \leq C_0 / P$ and $C_O \geq 6C_G / 7$ from which position we are not able to extricate easily. If there are better proofs, the author would be more than glad to hear them.

The bounds derived by Theorem 3 is still a posteriori since we have to compute the Graham's schedule and find the last job scheduled. To derive bounds a priori (which makes theoretical predictions about the complexity easier), we introduce the concept of Possible Last Job (PLJ) and illustrate with an example. We can compute PLJ and derive a priori bounds for the Graham's schedule when compared to the optimal one.

Let $K = \{ t_1, t_2, \ldots, t_n \}$ be a given set of jobs whose times are arranged in the descending order (note to minimize notation, we implicitly associate job $J_i$ with time $t_i$ and often use $t_i$ to denote job $J_i$ as well). Thus we have $t_1 \geq t_2 \geq \ldots \geq t_n$ and the Possible Last Job characterizes the job that can finish the last in Graham's schedule. If there is a job $t_i$ for $i < n$ such that we have:

$$t_i \geq (t_{i+1} + t_{i+2} \ldots + t_n), i < n \qquad (13)$$

we call such a job as dominant. PLJ is the index of the largest of the dominant job. As an example if we have $\{12, 5, 3, 2, 1\}$, we see $J_3$ with processing time 3 is dominant since $3 \geq (2 + 1)$, $J_2$ is not dominant but $J_1$ is since $12 \geq (5 + 3 + 2 + 1) = 11$. Note job $J_{n-1}$ is always dominant since $t_{n-1} \geq t_n$. In this case the PLJ is 1 since $J_1$ is the largest dominant job. The concept behind PLJ is that in a Graham's schedule PLJ *can* be the job that finishes the last but it is certainly possible other jobs whose indices are greater can finish last. In this example $J_1$ does finish last but if we have tasks $\{7, 5, 3, 3, 1\}$, then the only dominant job is the trivial $J_4$ and hence the PLJ is 4. In this case $J_4$ does not finish the last but if we consider tasks $\{7, 6, 3, 3, 2\}$, $J_5$ finishes the last. Thus we conclude:

**Lemma 1:** The index of the last job processed in a Graham's schedule is always greater than or equal to the PLJ for the given set of jobs. Also the PLJ can be computed with a complexity of $o(n^2)$ operations.

Let $P = \lceil PLJ / 2 \rceil$, i.e., P is the least integer $\geq PLJ / 2$. Since we have $L \geq PLJ$, we can follow a similar argument as in Theorem 3 to establish:

**Theorem 4:** $C_G / C_O \leq (Q + 1) / Q - 1 / 2Q$ where $Q = 24P^3 / (7 + 12P + 24P^2)$

Note the bounds derived in Theorem 4 can be computed without explicitly computing the underlying Graham's schedule. A numerical simulation is conducted to see how the various bounds compare and the results are summarized in Table 1. We considered 15, 20 and 25 jobs whose processing times are randomly chosen from [1, 32000] using a uniform distribution. For each set the simulation was run around 100 times and we computed the average completion time ratio (AC), the maximum completion time ratio (MC), the bound computed using Theorem 3 (BM), the bound computed using Theorem 4 (BP), and the bound BL computed using Coffman and Sethi (Theorem 2).

TABLE 1

| Jobs | AC | MC | BM | BP | BL |
|---|---|---|---|---|---|
| 15 | 1.007 | 1.045 | 1.080 | 1.086 | 1.068 |
| 20 | 1.004 | 1.020 | 1.055 | 1.061 | 1.051 |
| 25 | 1.002 | 1.016 | 1.044 | 1.046 | 1.040 |

Thus we see the bounds derived using Theorems 3 and 4 are comparable to the bounds derived by Coffman and Sethi and they get better with the increasing number of jobs.

We conclude the paper by presenting asymptotic bounds to the 2PS which are equally applicable to NPP. For a given set of N tasks, we divide them into two mutually exclusively subsets consisting of jobs { $t_1, t_2, \ldots, t_n$ } and { $t_{n+1}, t_{n+2}, \ldots, t_{n+m}$ } where n is the PLJ for the given set and n + m = N. Let $G_n$ and $G_{n+m}$ be the Graham's completion times for the sets { $t_1, t_2, \ldots, t_n$ } and { $t_1, t_2, \ldots, t_n, t_{n+1}, t_{n+2}, \ldots, t_{n+m}$ } and $O_n$ and $O_{n+m}$ be their optimal completion times. We note since PLJ = n, we have $t_n \geq t_{n+1} + t_{n+2} + \ldots + t_{n+m}$.

We have $O_n \leq O_{n+m}$ and $nt_1 \geq O_n$. Also $G_{n+m} \leq G_n + t_n$. Hence:

$$G_{n+m} / O_{n+m} \leq G_{n+m} / O_n \leq G_n / O_n + t_n / O_n \leq G_n / O_n + t_n / nt_1 \qquad (14)$$

Assuming $\delta = t_n / t_1$ and using Theorem 4 we have:

Theorem 5: For a set of N jobs with PLJ = n and $\delta = t_n / t_1$ is the ratio of the processing times of the dominant job to the largest job, as N becomes large we have:

$$C_G / C_O \sim 1 + 1/n + \delta/n \qquad (15)$$

Note that a similar type of result appears in Ibarra and Chen [7] (but their attribution of a similar result to Coffman and Sethi because of the referee's comments seems questionable to me).